\documentclass{elsarticle}

\usepackage{lineno,hyperref}

\begin{document}

\begin{frontmatter}

\title{Cosmology in One Dimension: The Two Component Universe}


\author{Yui Shiozawa}
\author{Bruce N. Miller\corref{mycorrespondingauthor}}
\address{Texas Christian University 2800 S. Univerisity Dr., Fort Worth}
\cortext[mycorrespondingauthor]{Corresponding author}
\ead{b.miller@tcu.edu}


\begin{abstract}
We investigate structure formation in a one dimensional model of a matter-dominated universe using a quasi-newtonian formulation. In addition to dark matter, luminous matter is introduced to examine the potential bias in the distributions. We use multifractal analysis techniques to identify structures, including clusters and voids. Both dark matter and luminous matter exhibit fractal geometry as the universe evolves over a finite range. We present the results for the generalized dimensions computed on various scales for each matter distribution. We compare and contrast the fractal dimensions of two types of matter for the first time and show how dynamical considerations cause them to differ. \end{abstract}

\begin{keyword}
Cosmology \sep Large-scale structure \sep Multifractal \sep Generalized Dimension \sep One-dimensional  \sep Dark Matter
\end{keyword}

\end{frontmatter}


\section{Introduction}

According to galaxy surveys, the universe appears to have large-scale, hierarchical structures up to a certain scale \cite{steidel1998large, tully2014laniakea}. Gravitationally bound collections of luminous galaxies are grouped into clusters and super clusters separated by large voids. As the cosmological principle states that the universe is homogeneous and isotropic at large scales, considerable effort has been made to study the scale at which the universe becomes homogenous \cite{springel2005simulations}. In order to understand the structure of the universe and its associated scale, we need to understand the distribution of dark matter, as it comprises the majority of the matter content of the universe \cite{bertone2005particle}. Since the visible galaxies are the only observational tracers, it is important to compare the evolution of each type of matter in a single model and investigate the possible bias against the distribution of dark matter. Therefore, we need to examine in what form the presence of dissipative baryonic matter affects the overall distributions.

Fractal analysis has proven to be a powerful tool in identifying scale-dependent structures as well as in quantifying their textures \cite{hogg2005cosmic, martins2006fractal}. It has been successfully applied to find the homogeneity scale both in large scale galaxy surveys and simulations \cite{bagla2008fractal}.  Unlike three-dimensional simulations, a one-dimensional model permits analytical solutions which allow us to maintain fractal fine structures. Therefore, with a one-dimensional  model, we can study the non-linear dynamics of the expansion with confidence. In the past, one dimensional models have shown robust scaling ranges, evidence of fractal-like structures \cite{miller2010cosmology, benhaiem2014self}. Accordingly, it is of wide interest to examine how a one-dimensional model universe with two matter components evolves over time. Our results can shed light on the large-scale structure of the actual universe. In particular, to gain information about both high and low density regions of the matter distribution, we employed mass-oriented methods which allow us to investigate multifractal spectra $D_q$, including the negative range of the index ${q}$ where popular size-oriented methods are known to have difficulty in producing accurate estimates \cite{Riedi95}.
	
\section{One-dimensional Model} The one-dimensional model was first formulated by Rouet and Feix \cite{Rouet1}. Other researchers also have worked on one dimensional models with different coefficients. For details, see the review paper by Miller et al.. \cite{miller2010cosmology} as well as \cite{miller2015cosmology} and \cite{benhaiem2014self} for more recent work. In this work, we extend the model to include luminous matter in addition to dark matter. To accomplish this we adopted a simple collision scheme such that luminous matter particles lose energy in interaction with each other. In contrast with dark matter, additional short range forces in luminous matter result in energy loss via radiation, turbulence, etc. Here we lump these effects into an effective inelastic collision between the ``luminous" particles.
	In formulating a one-dimensional model, we embed a set of infinitely large, two-dimensional, parallel sheets of mass with a density $m$ perpendicular to the configuration space. Since the fields generated by the sheets of mass are independent of their position and are parallel to the configuration space, we can confine their motions to an effectively one-dimensional space. Therefore, we represent a sheet by a particle which moves along the configuration space. In order to reduce boundary effects, we customarily employ periodic boundary conditions which take into account the infinite number of replicas of the mass sheets contained in the original interval ${[-L, L)}$. While the potential from the infinite number of masses diverges, we can benefit from a technique called Ewald summation. Using this technique, we can isolate the potential which gives rise to the motion of particles by subtracting the background potential \cite{miller2010ewald}. In this way, it can be shown that the total field ${E(\chi)}$ from the number of particles ${N}$ in the original interval ${[-L, L)}$ is
\begin{equation} \label{eq: boundary}
E(\chi)= \left[ \frac{N}{L} (\chi-\chi_c) + \frac{1}{2} \left( N_R(\chi)-N_L(\chi) \right) \right]
\end{equation}
where ${\chi_c}$ is the center of mass of the system and ${N_R(L)}$ is the number of particles to the right (left) of the position $\chi$ within the original interval  \cite{miller2010ewald}. Following standard practice, we set up a dynamical equation using Newtonian mechanics with co-moving coordinates. During the matter-dominated universe in which structure formation takes place, the universe expands roughly by a scale factor ${a(t) \propto (t/t_0)^{2/3}}$ \cite{peebles1993principles} for some time unit ${t}$ where the initial time ${t_0}$ may be set to the epoch of recombination, i.e. the beginning of the matter-dominated universe. We introduce a co-moving coordinate ${\chi}$ such that the apparent length is kept fixed. The co-moving coordinate is related to the original coordinate ${r}$ by ${r= a(t) \chi}$. Due to this transformation, we can rewrite the field equation in terms of the co-moving coordinate. By introducing a logarithmic time scale ${T}$ and an appropriate time unit, we obtain
\begin{equation}\label{eq: RF}
\frac{d^2\chi}{dT^2}+\frac{1}{\sqrt{2}}\frac{d\chi}{dT}-\chi=E(\chi,T).
\end{equation}
This is the signature equation of motion in the RF model, named after Rouet and Feix, and its formulation is fully discussed in their work \cite{Rouet1}.
With the ``friction'' coefficient being ${\frac{1}{\sqrt{2}}}$ in the RF model, we can analytically obtain the crossing time between two particles by solving cubic equations. Thus we can write an event-driven algorithm and minimize the unknown effects often brought in by numerical approximations. In this work, we extend the previous model by introducing luminous matter. In the simulation, luminous matter and dark matter behave identically except at the crossings. When two luminous matter particles approach, they ``collide'' and lose energy in interaction with each other. We set a velocity-dependent collision coefficient $\kappa$ analogous to a restitution coefficient. The velocity dependence is given by $\kappa=\exp \left( -c|v_1-v_2|^{3/5} \right)$ where $v_1$ and $v_2$ represent the velocities of two colliding particles. The coefficient $c$ was chosen arbitrarily in the simulation so that the trajectories of luminous matter particles are substantially different from dark matter particles without forcing them to collapse too fast. The luminous particles lose more energy when the velocity difference between the two is large. Initially, the particles are placed near the equilibrium positions which are separated equally in the configuration space.

\section{Initial Conditions} The primordial potential fluctuation is chosen to replicate the scale-invariant Harrison-Zel'dovich spectrum \cite{Harrison1970}. In a three-dimensional universe, the spectral index ${n}$ for the power spectrum is unity which roughly agrees with the estimate from observations \cite{dunkley2009five}. In the one-dimensional case, the spectral index ${n}$ needs to be three to insure that potential fluctuations are invariant of scale. We randomly assigned initial positions so that the fluctuations around the equilibrium positions follow these statistics. Based on observational estimates, the dark matter to luminous matter ratio is fixed to 4:1 \cite{bertone2005particle}. Accordingly one fifth of the total particles are selected using a random process and designated as luminous matter. The results presented in this work were performed with the total number of particles ${N = 100,000}$. For simplicity, we chose units such that the original interval length ${2L}$ is equal to the number of particles ${N}$.
\begin{figure}
\begin{center}
{\includegraphics[width = \textwidth]{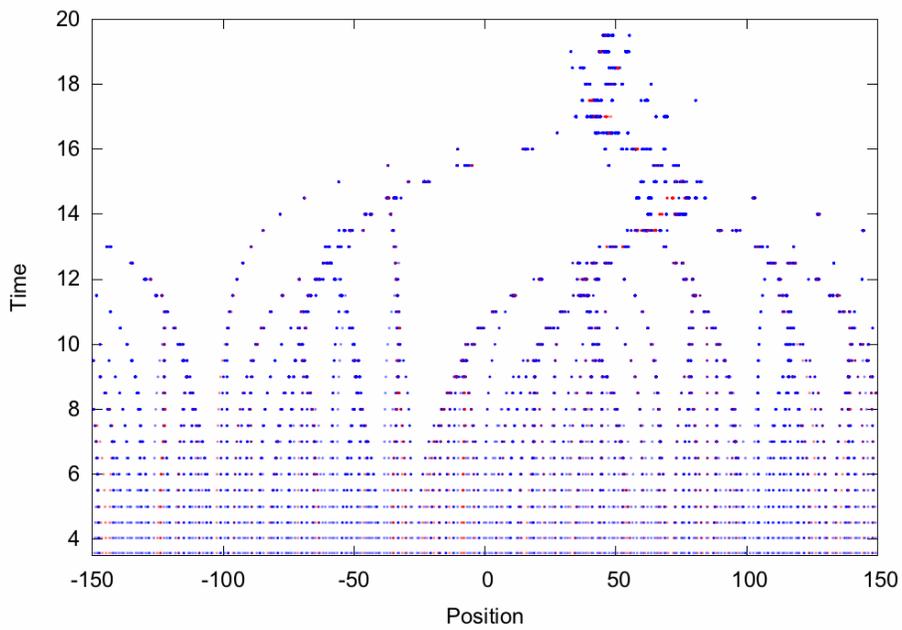}}
\caption{This figure shows how matter coalesces together to form a cluster at the end of the simulation. The initial distribution of particles is nearly uniform. The simulation was perfumed with $N=300$ for illustrative purposes. The blue dots represent dark matter and red dots luminous matter.  } \label{fig: evolution}
\end{center}
\end{figure}

\begin{figure}
\begin{center}
{\includegraphics[width = \textwidth]{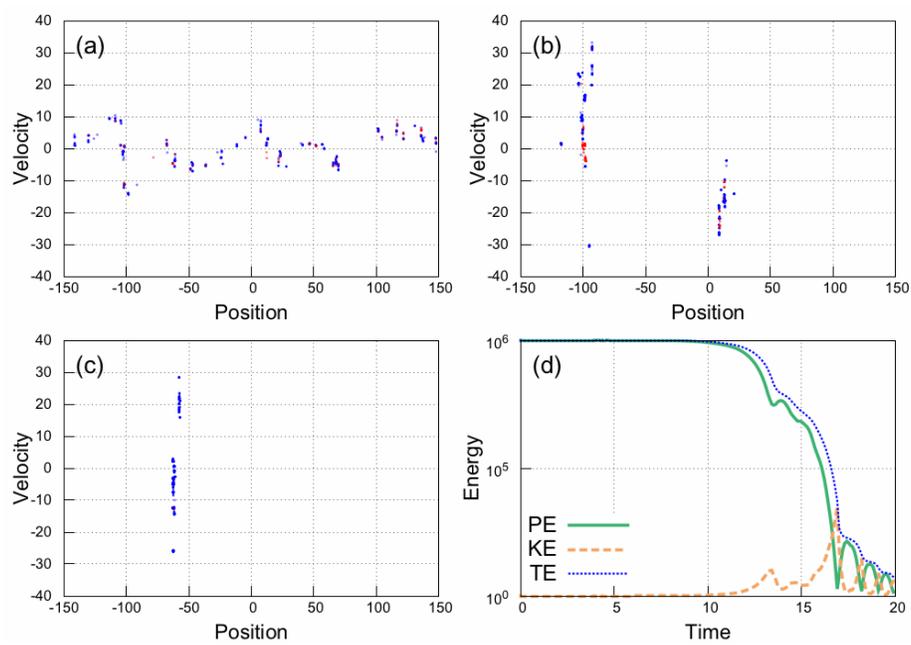}}
\caption{Figures (a), (b) and (c) show the distribution of 300 matter particles in $\mu$-space at $T=10$, $15$ and $20$ respectively. The blue dots represent dark matter and red dots luminous matter. Figure (c) illustrates the energy of the system over time.} \label{fig: energy}
\end{center}
\end{figure}

	\section{Simulations} In Fig. \ref{fig: evolution}, we show how the distribution of matter in space evolves over time. For illustrative purposes, this simulation was performed with a smaller number of particles ${N= 300}$. The initial positions of the particles are shown at the bottom of the figure and, by looking towards the top, we see that the particles coalesce to form clusters. Viewing on line one can distinguish the the non-dissipative dark matter from the dissipative luminous matter by color. In Fig. \ref{fig: energy}, we extracted snapshots of the matter distribution at a few different times $T$ in $\mu$-space where the vertical axis represents the velocities of the particles and the horizontal axis the positions. With ${N=100,000}$, the system undergoes a similar evolution but on a more massive scale. With ${N=300}$, we observe that the nearly homogenous initial distribution evolves into a single cluster towards the end of the simulation. At the core of the cluster, luminous matter appears to be concentrated with dark matter forming a halo around it. We can also see how the energy of the system evolves in Fig. \ref{fig: energy}. Initially, potential energy dominates the system. As the total energy decreases due to the friction term in the equation of motion Eq.~(\ref{eq: RF}), the particles begin to pick up kinetic energy. While smaller clusters interact with each other and merge into a larger cluster, potential and kinetic energy exchange. The small, random fluctuations in total energy are due to the collisions between luminous particles.
	
\section{Fractal Analysis} By comparing the fractal dimension of the set with the dimension of the embedding space we can estimate the degree of inhomogeneity and complexity. In order to study the formation of the clusters and voids separately, in this work we used the generalized fractal dimensions, for which the well-known box-counting dimension is a special case. The generalized dimensions, or Renyi dimensions, are defined by \cite{falconer2003}:
\begin{equation} \label{eq: Renyi}
D_q=-\frac{1}{1-q}\lim_{\epsilon \to 0} \frac{\ln \sum_{i=1}^{N(\epsilon)} p_i^q}{\ln \epsilon} \end{equation}
where $N(\epsilon)$ is the number of covering sets with diameter $\epsilon$ required to cover a given mass distribution and $p_i$ is the probability associated with the $i$\textsuperscript{th} subset. Since the parameter ${q}$ can assume any real number, a spectrum of dimensions, instead of a single value, can be obtained for a given set.
	For a standard non-fractal set, the fractal dimension coincides with the dimension of the embedded space for all ${q}$. Therefore, the values of ${D_q}$ away from the dimension of the embedded space represent the inhomogeneity of the set. Furthermore, the generalized dimensions can separately analyze subsets with different degrees of density in a set. The positive and negative values of ${q}$ correspond to the dense and sparse regions of the set, respectively. In our model, the regions with a high density of matter particles can be interpreted as clusters. Likewise, the regions with a low density of matter particles may be called voids. Thus, with multifractal analysis, we can separate the evolution of clusters from the evolution of voids.
	While Eq.~(\ref{eq: Renyi}) provides the standard definition for the generalized dimensions, in practice, various numerical methods can be applied to estimate them. Among all, the box-counting method may be the most widely-used numerical approach. However, this method, as well as other methods using partitions of equal size, are known to have difficulty estimating the generalized dimensions for the negative range of ${q}$ \cite{Riedi95}. Instead, we applied mass-oriented methods, all of which may be considered variants of the method originally proposed by van de Water and Schram \cite{Water88}. In general, mass-oriented methods have advantages over size-oriented methods on low-density regions and are therefore more suitable for studying void structure formation \cite{shiozawa2014}.

	Mass-oriented methods are based on the statistics of the ${k}$\textsuperscript{th} nearest distances from reference points sampled from a given set. For each reference particle, we first determine the ${k}$\textsuperscript{th} closest particle and then compute its distance from the reference point. According to van de Water and Schram, a series of exponents ${D(\gamma)}$ exists for each $\gamma$ such that:
\begin{equation} \label{eq: k-neighbor}
\left<\Delta^{\gamma}(k,n) \right>^{1/\gamma}  \cong n^{-1/D(\gamma)} \left[\alpha D(\gamma) \frac{\Gamma(k+\gamma/D(\gamma))}{\Gamma(k)} \right]^{1/\gamma}
\end{equation}
where $\alpha$ can depend weakly on $k$ but is independent of $\gamma$ \cite{Water88}. As the notation suggests, in the mass oriented methods, the Dimension Function ${D(\gamma)}$ plays a similar role as the generalized dimensions ${D_q}$ and may be considered alternative multifractal dimensions. In fact, the following implicit equation provides the relation between the two \cite{Water88}: $D[\gamma = (1-q)D_q] = D_q$.

	In estimating the dimension function ${D(\gamma)}$, we can fix the value of ${k}$ and let the number of reference points $n$ increase or vice versa. The exponent ${D(\gamma)}$ can then be found as the slope of the best-fit line in a scaling range in a log-log plot. In an ideal situation, the values of the weighted sum in Eq. \ref{eq: k-neighbor} all line up in a log-log plot with a single associated slope. In practice, a scaling range is finite and identifying it can play a crucial role. When the value of ${k}$ is fixed with some small integer, the method is often referred to as the near neighbor method in the literature with a special case, called the nearest neighbor method, for ${k=1}$ \cite{Badii85}. The near-neighbor method is also known to be problematic when estimating the generalized dimensions in the positive range of ${q}$, or equivalently, the Dimension Function in the negative range of $\gamma$  \cite{shiozawa2014}. The theoretical range where the singularities exist is for ${k < \frac{\gamma}{D(\gamma)}}$. For this reason, we kept the value of ${k > 5}$ so we expect to obtain reasonable results for the multifractal spectra in both positive and negative ranges. In addition to the near-neighbor method, Broggi has studied the fixed-${k}$ approach with large ${k}$ values  \cite{broggi88}. Here, following Broggi, we also consider values of ${k}$ up to 1000, i.e., a hundredth of the number of test particles in the system. If a set is a fractal on all scales, the associated dimensions should be independent of the choice of the value of ${k}$. In dynamics, however, a simulated set often exhibits scale-dependent properties, meaning that the associated dimensions vary with the value of ${k}$. 	
	Instead of fixing the value of ${k}$, we can also fix the number of the reference points and study how scaling arise when increasing the value of ${k}$. From the earlier studies on mass oriented methods, this ${k}$-neighbor approach is known to work on the entire range of the spectrum but the results are highly sensitive to the choice of a scaling range  \cite{shiozawa2014}. If a given set has scale-dependent properties, multiple scaling ranges may exist. Unlike the fixed-${k}$ approach, the scaling ranges may not exhibit clear-cut limits and correction factors need to be introduced in order to obtain better results. Nevertheless, the k-neighbor approach allows us to study the scale dependent properties simultaneously and provide a global perspective on a given set. For these reasons, we use this approach as a starting point and show how the matter distribution evolves globally without necessarily extracting the Dimension Functions.
\begin{figure}
\begin{center}
{\includegraphics[width = \textwidth]{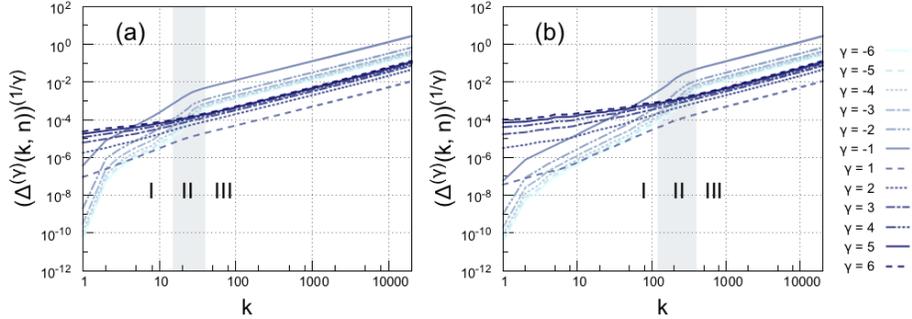}}
\caption{Figures (a) and (b) show the scaling obtained with the $k$ neighbor method at $T=14$ and $T=19$ respectively.  The shading indicates three different zones.}\label{fig: scaling}
\end{center}
\end{figure}
\section{Results} We present the results of the generalized dimensions ${D_q}$, computed with the fixed-${k}$ approach, with a selected set of ${k}$ values over time. However, to get an idea of how structure evolves in time, we first consider the $k$-neighbor approach. Although the scaling ranges for luminous matter do not coincide with dark matter, they share several essential features. Accordingly, we may subdivide each plot into three zones. For each $\gamma$, we see that each plot exhibits some scaling range over relatively small values of ${k}$ (zone I). This scaling range ends at some point and another scaling range (zone II) with a different slope appears. For large values of ${k}$, every plot seems to have identical slope values (zone III). At earlier times, this final zone starts at smaller ${k}$ and begins to recede as the universe evolves. There is approximately a single slope associated with this zone. Therefore, zone III corresponds to the homogeneous part of the matter distribution. It also follows that this one dimensional universe is homogenous at almost all scales at the beginning: At later times, it is only homogenous at large scales. The first two zones have slopes different from 1 so they represent inhomogeneity in the distribution. The first scaling range grows over time. By comparing the evolution in the ${k}$\textsuperscript{th} neighbor scaling and the dynamics in the $\mu$-space diagrams, the first scaling range seems to represent the clusters and the second scaling range is the transient state between the homogenous distribution and the cluster distribution. This transient regime increases in ${k}$ with time, but the scaling range does not expand substantially. Rather, it shifts to the right which also explains why it is a transient state.
	The threshold values between the homogenous and transient regimes depend on the type of matter. This suggests that the total distribution of matter is a union of two distinct distributions and they need to be investigated separately. Dark matter has threshold values approximately 4.5 times higher than luminous matter for each corresponding index $\gamma$. This ratio is close to the dark matter to luminous matter ratio set at the beginning of this simulation. Thus each matter distribution has a similar homogenous scale given that the matter ratio is maintained at the same scale.
	
\begin{figure}
\begin{center}
{\includegraphics[width = \textwidth]{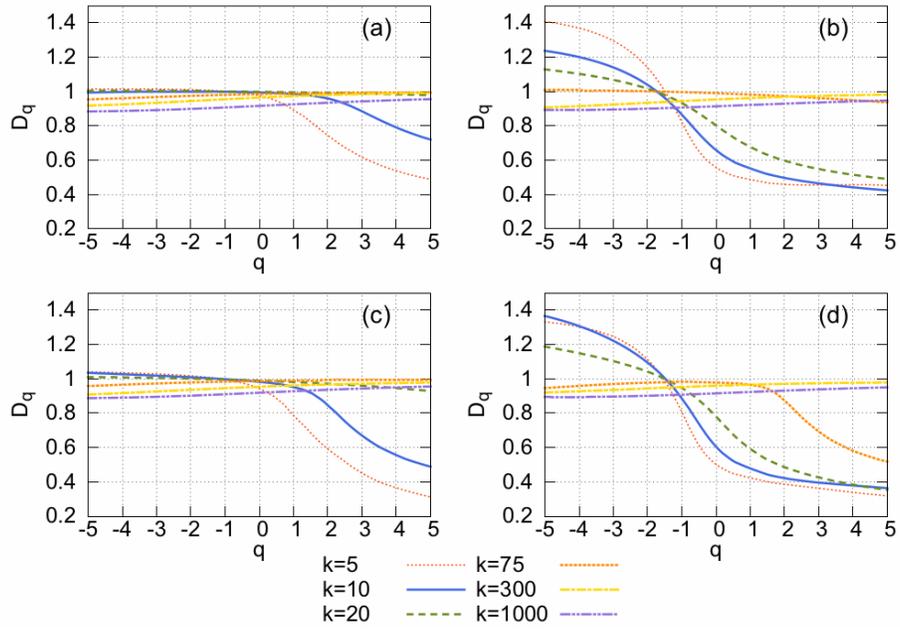}}
\caption{ Figures (a) and (b) show the generalized dimension $D_q$ computed with the fixed-$k$ approach at a selected set of $k$ values for dark matter at $T=14$ and $19$, respectively. Figures (c) and (d) show $D_q$ for luminous matter at $T = 14$ and $19$.} \label{fig: Dq}
\end{center}
\end{figure}
	We applied the fixed-${k}$ approach to estimate the multifractal spectrum of the system. We chose a set of ${k}$ values so that, at the end of the simulation, the results can be associated with the clusters, the transient state and the homogeneous range. At the beginning, the generalized dimensions are uniform and equal to unity for all values of ${k}$. As time progresses, the positive range of ${q}$ starts to diverge from 1 at ${k=5}$ as shown in Fig. \ref{fig: Dq}. In this range, the generalized dimensions eventually reach the lower limit at ${D_q=0.4}$. Once this limit is reached, we observe no further development. This is consistent with the well-known stable clustering hypothesis which states that the mean separation of particles remains constant on sufficiently small scales \cite{Jing2001}. The generalized dimensions at higher values of ${k}$ undergo a similar development but the larger the value of ${k}$, the longer it takes to begin the process. Likewise, the generalized dimensions in the negative range of ${q}$ begin to diverge from 1 but at later times, compared to the positive range of ${q}$. The spectrum is less flat in the negative range, and so the estimates appear to suffer from numerical difficulties. Nevertheless, the results clearly show how the generalized dimensions increase in this range over time and its spectrum with smaller ${k}$ values begin developing at earlier times. Luminous matter also shows a similar development but there are a few quantitative differences. The lower limit for the generalized dimensions is smaller. The positive range shows signs of inhomogeneity at earlier times for a given ${k}$. The spectrum in the negative range generally follows dark matter.
	
\begin{figure}
\begin{center}
{\includegraphics[width = \textwidth]{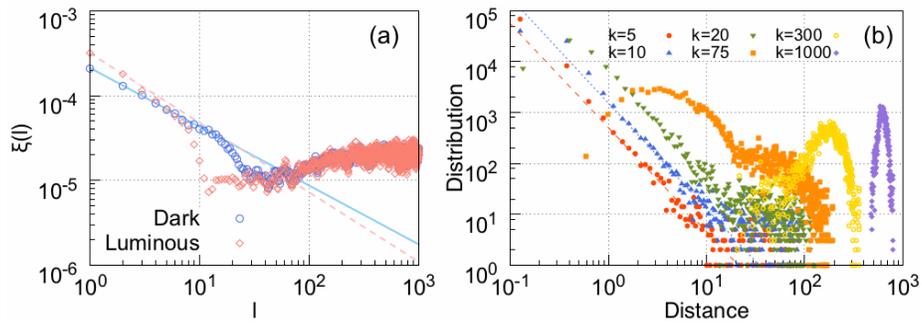}}
\caption{ The correlation function was computed to obtain an estimate for $D_2 = 0.31$ for dark matter and $D_2 = 0.18$ for Lu-minous Matter. For each $k$\textsuperscript{th} neighbor distance, the distributions were plotted for the selected set of time T.} \label{fig: correlation}
\end{center}
\end{figure}
	
	To verify these claims, we conducted several alternative numerical analyses. As the results obtained from a single numerical method are often not reliable, we used the correlation function to estimate the correlation dimension of each matter distribution. As shown in Fig. \ref{fig: correlation} the results for $D_2$ are slightly lower than the earlier results obtained from the fixed-$k$ approach. However, since luminous matter consistently exhibits higher inhomogeneity, we conclude that the presence of the short range force creates a distinct distribution texture from dark matter within a cluster. In order to map the orders of $k$ to the size of structures in the configuration space, we studied the probability distribution of the $k$\textsuperscript{th} nearest neighbor distance for each type of matter. As shown in Fig.\ref{fig: correlation}, the distribution shifts to the right with increasing value of $k$, confirming the previous assertion that the fractal dimensions depend on their associated scale and the hierarchical structure forms from bottom-up. Since a scaling range clearly exists for the orders of k which were previously associated with clusters, we computed a straight line best-fitted to each scaling range and found its $x$-intercept. We then compared the $x$-intercept obtained from each type of matter for a given $k$. As expected, the larger the $k$ value becomes, the larger their associated $x$-intercept we obtained.
	
	\section{Conclusions} In summary, we employed mass-oriented methods to find that both dark matter and luminous matter distributions reveal rich fractal structures. At the beginning of the simulation, the model universe is homogenous on all scales. The inhomogeneity continuously grows and the scale at which the distribution becomes homogenous expands over time. Each type of matter follows a similar evolution with significant quantitative differences. Inside clusters, luminous matter is concentrated at the core, giving rise to lower fractal dimensions in the positive range of ${q}$. Due to the energy loss during the collisions, luminous matter starts to coalesce first. On the contrary, the void regions show no significant difference between the two types of matter as the long range force is primarily responsible for void formation. The structures continues to increase in size and the fractal patterns persist over time. The difference in fractal dimensions within clusters is a manifestation of the bias of the luminous matter distribution against the dark matter distribution. This phenomena will be investigated more completely in a longer work.
	
\section{acknowledgments}
We thank Dr. Jean-Louis Rouet and Dr. Igor Prokhorenkov for their insights and expertise. We also thank Dr. Benjamin Janesko and Mark O'Callaghan for assistance with the TCU cluster.
\section*{References}

\bibliography{reference}

\end{document}